\documentclass[sigconf]{acmart}
\usepackage{graphicx}
\usepackage{graphicx}
\usepackage{multirow}
\usepackage{booktabs}
\usepackage{amsmath}
\usepackage{makecell}
\usepackage{subcaption}
\usepackage{enumitem}

\setcopyright{none}
\AtBeginDocument{%
  }

\copyrightyear{2026}
\acmYear{2026}
\setcopyright{cc}
\setcctype{by}
\acmConference[SIGIR '26]{Proceedings of the 49th International ACM SIGIR Conference on Research and Development in Information Retrieval}{July 20--24, 2026}{Melbourne, VIC, Australia}
\acmBooktitle{Proceedings of the 49th International ACM SIGIR Conference on Research and Development in Information Retrieval (SIGIR '26), July 20--24, 2026, Melbourne, VIC, Australia}
\acmDOI{10.1145/3805712.3809557}
\acmISBN{979-8-4007-2599-9/2026/07}

\usepackage{verbatim}
\usepackage{graphicx} 
\usepackage{float} 
\usepackage{multirow}
\usepackage{fancyhdr}
\usepackage{enumitem}
\usepackage{xspace}
\usepackage{colortbl}
\usepackage[table,xcdraw]{xcolor}

\newcommand{\eg}{\emph{e.g., }}

\begin{document}


\title{Dual-Diffusional Generative Fashion Recommendation
}


\author{Mingzhe Yu}
\email{mingzhe.yu.2025@phdcs.smu.edu.sg}
\affiliation{%
  \institution{Singapore Management University}
  \country{Singapore} 
}

\author{Lei Wu}
\email{i_lily@sdu.edu.cn}
\affiliation{%
  \institution{Shandong University}
  \city{Jinan}
  \country{China}
}

\author{Qianru Sun}
\email{qianrusun@smu.edu.sg}
\affiliation{%
  \institution{Singapore Management University}
  \country{Singapore}
}

\author{Yunshan Ma}
\email{ysma@smu.edu.sg}
\authornote{Corresponding author.}
\affiliation{%
  \institution{Singapore Management University}
  \country{Singapore}
}

\renewcommand{\shortauthors}{Mingzhe Yu, Lei Wu, Qianru Sun, and Yunshan Ma}


\begin{abstract}
Personalized generative recommender systems have emerged as a promising solution for fashion recommendation. However, existing methods primarily rely on implicit visual embeddings from historical interactions, which often contain preference-irrelevant information and result in insufficient user behavior modeling. Moreover, these models typically generate only item images, providing limited interpretability.
To address these limitations, we propose DualFashion, a Dual-Diffusional Generative Fashion Recommendation Architecture that jointly models image and text modalities for personalized and explainable recommendation. DualFashion adopts a dual-diffusion Transformer with image and text branches, where structured attribute-level captions and visual outfit information are jointly used as conditioning signals to model user behavior. The proposed architecture produces both fashion item images and textual descriptions, ensuring visual compatibility while providing explicit semantic interpretability.
Furthermore, we introduce a text-augmented fine-tuning strategy that enhances generation diversity and enables effective cross-modal knowledge transfer without incurring heavy computational costs. Extensive experiments on iFashion and Polyvore-U across Personalized Fill-in-the-Blank and Generative Outfit Recommendation tasks demonstrate that DualFashion achieves strong performance in behavior modeling, interpretability, and efficiency compared to state-of-the-art methods.
Our code and model checkpoints are available at \url{https://github.com/LinkMingzhe/DualFashion}.
\end{abstract}







\maketitle

\section{Introduction}

Recommender system has been widely applied on digital platforms to route personalized content to users catering to their personalized preference \cite{PMG, POG, liu2022elimrec, liu2026principled}.
Early research\cite{CAR, DGSR, CLHE, CIRP, Bias, RMBRec} primarily relies on ranking-based approach, which outputs a ranking list and retrieves items from existing catalogs.
Recently, with the remarkable advancement of generative AI, such as text, image, and music generative models, 
generative recommender systems\cite{survey_pgen, LLM4DSR}, which directly generate the content of items instead of retrieving them from existing catalogs, have garnered growing attention from both academia and industry communities.


\begin{figure}
  \includegraphics[width=0.47\textwidth]{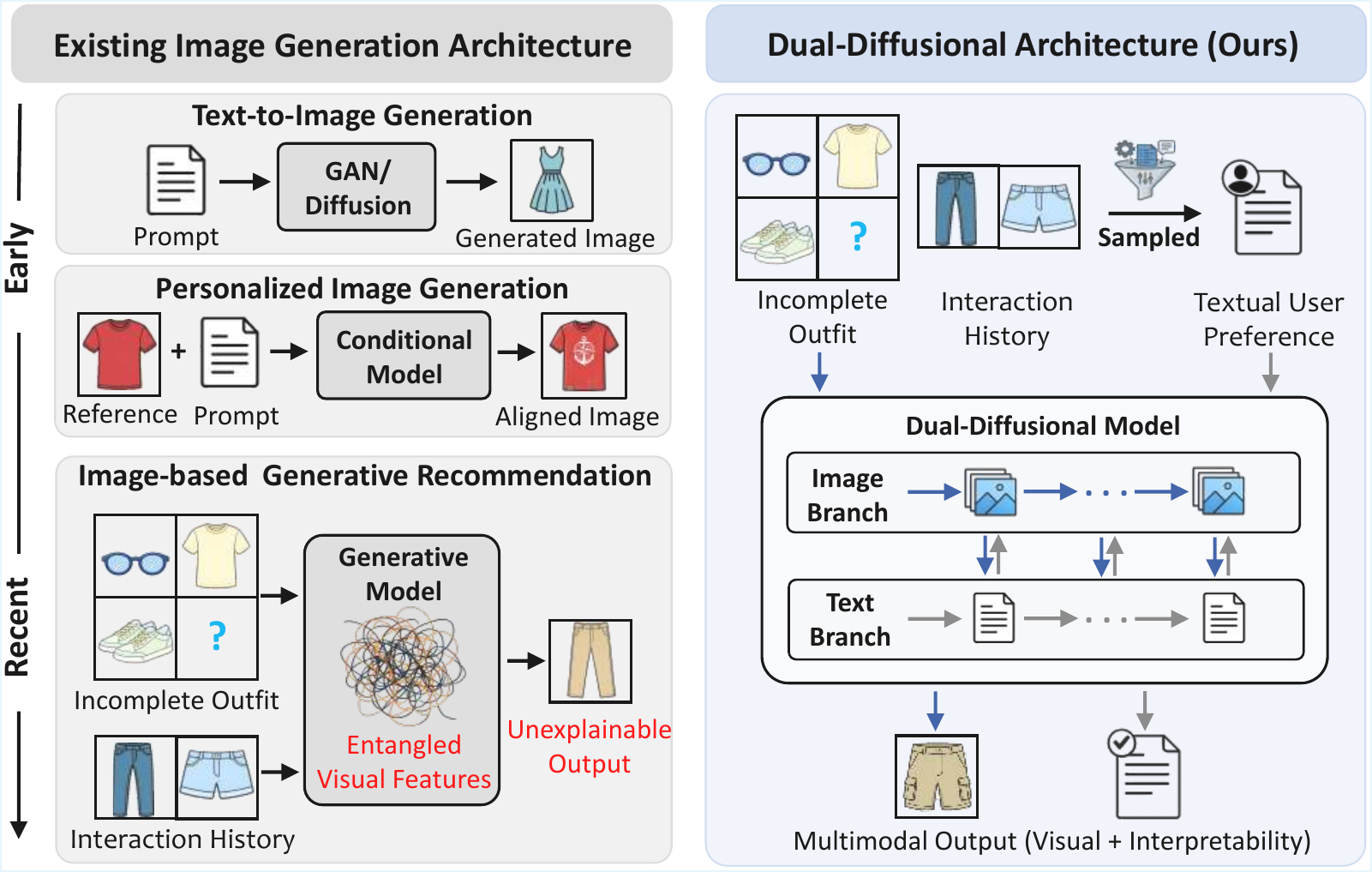}
  \setlength{\abovecaptionskip}{-0.2cm} 
  \setlength{\belowcaptionskip}{-0.2cm}
  \caption{Comparison between our dual-diffusional architecture and existing fashion image generation architectures.}
  \label{Intro-frontPage}
\end{figure}


With continuous advances in image generation and generative recommendation, image-based generative recommendation has achieved significant progress,
yet existing methods are still limited to the image modality.
Early research can be traced back to text-to-image models that aim to synthesize high-quality images.
GANs \cite{GANs, cGANs} formulate image synthesis as a minimax optimization problem between generator and discriminator.
Diffusion-based models \cite{SD,SDv3} denoising images in latent space, enabling high-fidelity image synthesis from textual prompts.
Subsequent works, such as ControlNet \cite{controlnet}, further introduce conditional information to enable more fine-grained control.
Personalized image generation \cite{DreamBooth, Textual_Inversion} further extends conditional generation by introducing user-provided reference image.
More recently, researchers \cite{Difashion, PMG} integrate image generation into generative recommendation, which consider user-item interaction as preference and generate images that user might find interesting.
However, these methods generate outputs solely in the image modality and do not fully exploit textual information, which limits the model expressive capacity.

In the fashion domain, most image-based generative recommendation methods are built upon image-centric architectures, where both user modeling and item generation are dominated by visual signals. While effective for visual synthesis, this architectural choice gives rise to two fundamental limitations. First, user behavior modeling remains insufficient. Existing generative approaches~\cite{Difashion, FashionDPO} predominantly rely on interacted item images as unimodal inputs to capture fine-grained preferences. However, user–item interactions are inherently sparse, and visual data often contain redundant or preference-irrelevant details, making reliable preference extraction difficult. In contrast, textual modality can abstract away low-level visual noise and encode high-level semantic intent, a property that has long been exploited in traditional recommendation models for preference modeling.
Second, explainability is largely absent. Recent advances, from GAN-based models~\cite{OutfitGAN} to diffusion models~\cite{Difashion, FashionDPO} and large multimodal models~\cite{Pigeon}, continue to treat image generation as the sole output, providing little explicit rationale for why a generated item complements an incomplete outfit and fit the user's preference. Prior efforts attempt to mitigate these issues through disentangled behavior modeling~\cite{DiscRec, DGCF} or by incorporating textual information as auxiliary components~\cite{ERRA}. However, such solutions are typically add-on patches to image-centric backbones, increasing architectural complexity without fundamentally addressing the problem. We argue that both insufficient behavior modeling and lack of explainability share a common root cause: the restrictive image-centric architectural design. Consequently, resolving these limitations requires a paradigm shift toward multimodal architectures that natively support image–text inputs and jointly generate visual and textual outputs.




Implementing this idea, we propose a Dual-Diffusional Generative Fashion Recommendation Model Architecture (DualFashion), which uses multimodal input to comprehensively model user behavior and provide multimodal output as explainable fashion item recommendation.
1) We extract the fashion items' captions in attribute level and construct the user preference with weighted sampling.
Based on the dual-diffusion architecture, we disentangle use preference in text branch and outfit information in image branch, enabling effective modeling of user behavior by jointly leveraging both visual and textual information.
2) At the output level, our architecture produces visual images as well as textual descriptions for explicit explanations.
To be noted, visual and textual outputs are jointly optimized during training, allowing the two modalities to mutually reinforce each other.
Moreover, the dual-diffusion architecture facilitates effective knowledge transfer between modalities.
We augment captions with additional samples that satisfy both matching and personalization requirements, enabling more efficient training while enhancing generation diversity.



We conduct experiments on two established datasets, iFashion~\cite{POG} and Polyvore-U~\cite{Polyvore}, covering two standard tasks: Personalized Fill-in-the-Blank (PFITB) and Generative Outfit Recommendation (GOR)~\cite{Difashion}. 
By jointly modeling image and text modalities within a dual-diffusion Transformer and adopting a multi-stage training strategy, our approach achieves both strong visual compatibility and explicit textual interpretability.
In addition, the text-augmented fine-tuning enhances diversity in both textual and visual outputs.
The main contributions of this work are summarized as:

\begin{itemize}[leftmargin=0.5cm, itemindent=0cm]
\item We identify that insufficient user behavior modeling and limited explainability in existing generative fashion recommendation models stem from architectural design, and propose a dual-diffusional architecture to address these issues.



\item To the best of our knowledge, we are the first to introduce the dual-diffusion Transformer that jointly optimizes image and text diffusion processes for generative fashion recommendation.

\item Extensive experiments on two established datasets (iFashion and Polyvore-U) and two standard tasks (PFITB and GOR) demonstrate the effectiveness of our proposed architecture. Both the source code and datasets are open-sourced.


\end{itemize}

\section{Related Work}

\subsection{Image Generation}

Image generation aims to synthesize high-fidelity and diverse visual samples. In computer vision domain, the image generation has progressed from GAN-based approaches \cite{stylegan,bigGAN} to Diffusion-based methods \cite{DDPM, DDIM, DiT, LCM}.
With cGAN \cite{cGANs} and ControlNet \cite{controlnet} introduces condition into image generation process, personalized image generation has achieved significant advancements.
DreamBooth \cite{DreamBooth} enables instance-level personalization by fine-tuning generative models with a small number of subject images.
In parallel, researchers in the fashion domain have actively adopted these advancements for fashion image generation, aiming to create high-resolution, design-rich clothing items that satisfy personalized needs, such as virtual try-on, personalized fashion image generation, outfit generation.
In virtual try-on task, StableVTON \cite{stableVTON} learns semantic correspondences between clothing and the human body within the latent space of a pre-trained diffusion model.
For personalized fashion image generation, HMaVTON \cite{HMaVTON} generates personalized and matching fashion items for users.
In outfit generation, backbone architectures have evolved from GAN-based models to diffusion-based models and, further to multimodal models.
OutfitGAN \cite{OutfitGAN} learns compatible fashion outfits via adversarial training.
DiFashion \cite{Difashion} utilizes a diffusion-based architecture to generate fashion items for outfit completion or matching outfits. 
FashionDPO \cite{FashionDPO} employs a reinforcement learning strategy to fine-tune the outfit generation model via Direct Preference Optimization \cite{D3PO}.
The multimodal model Pigeon \cite{Pigeon} design multimodal instructions and predict the personalized image.
However, as generative capacity increases, model parameters continue to grow, making effective model training and fine-tuning an important research direction.

\subsection{Generative Recommendation}

Due to the rapid emergence of generative models, an increasing number of recommendation studies adopt generative paradigms, such as LLMs \cite{iEvaLM, wang2026mllmrec, 2025IRLLRec, MSL, LLM4DSR}, semantic IDs \cite{RSGR}, and image generation models \cite{GRModels-survey}.
In this paper, we focus on the image-based generative recommendation, which typically employs generative models \cite{GANs, SDv3} to synthesize high-fidelity visual content.
Unlike the ranking-based methods, these approaches can generate novel items beyond the fixed candidate dataset.
Existing works can be categorized into two streams based on their generative backbones:
1) GAN-based Approaches: DVBPR \cite{DVBPR} introduces a visually-aware fashion recommendation paradigm that integrates generative image models to jointly capture user preference modeling and controllable fashion item design.
While effective in capturing local patterns, GAN-based methods often suffer from training instability and mode collapse, limiting the diversity of recommended results.
2) Diffusion-based Approaches: By formulating recommendation as a reverse denoising process, these models can synthesize high-fidelity images from Gaussian noise conditioned on user behavior.
PMG \cite{PMG} leverages LLMs to model user preferences and generate visual outputs.
However, existing model architectures only output recommendation item images, failing to provide interpretability.
In our study, we introduce multimodal input and output architecture to produce visual image and textual description, which ensures the recommendation is not only visually compatible but also semantically interpretable.

\section{Preliminary}
We introduce the task definition, followed by the formulation of image and text generation.

\textbf{Problem Formulation.}
In the fashion domain, researchers \cite{Difashion, FashionDPO} define the image-based generative fashion recommendation tasks as: 
1) Personalized Fill-in-the-Blank (PFITB) - generating one matching item to complete the outfit, 
and 2) Generative Outfit Recommendation (GOR) - building one completed outfit.
Based on the user information $\boldsymbol{u}$, item information $\boldsymbol{i}$, these tasks aim to generate complete outfit $\boldsymbol{O} = \{\boldsymbol{i}_k\}_{k=1}^n$, where $\boldsymbol{O}$ denotes the complete outfit, $\boldsymbol{i}_k$ represents each compatible fashion item and $n$ is the number of fashion items included in the outfit.
1) PFITB: Given the incomplete outfit $\boldsymbol{O}^{'} = \boldsymbol{O} \backslash \{\boldsymbol{i}_0\}$, this task aims to generate compatible fashion items $\boldsymbol{i}_0$ based on users' interacted history. 
2) GOR: Without the incomplete outfit $\boldsymbol{O}^{'}$, this task generates a complete set of matching outfits $\boldsymbol{O}$ ($n$ items) for users. 

\textbf{Image Generation - Continuous Diffusion}
We define $\boldsymbol{z}_0$ as the latent representation of the recommended item image that matches the incomplete outfit $\boldsymbol{O}^{'}$. With the condition information user preference $\boldsymbol{p}$, matching condition $\boldsymbol{m}$ and task definition $\boldsymbol{d}$, our goal is to model the conditional image distribution: $p_\theta(\boldsymbol{z}_0 \mid \boldsymbol{p}, \boldsymbol{m}, \boldsymbol{d})$.
We adopt the continuous diffusion formulation with a forward process that gradually add noise to the clean latent $\boldsymbol{i}_0$.
Following flow-matching based continuous diffusion ~\cite{SDv3}, the forward process is defined as
\begin{equation}
    \boldsymbol{z}_t 
    \;=\; \alpha(t)\,\boldsymbol{z}_0 
    + \sigma(t)\,\boldsymbol{\epsilon},
    \quad t\in[0,1],\quad
    \boldsymbol{\epsilon}\sim\mathcal{N}(\mathbf{0},\mathbf{I}),
    \label{eq:forward}
\end{equation}
where $\alpha(t)$ and $\sigma(t)$ are scalar functions satisfying $\alpha(0)=1$, $\sigma(0)=0$ and $\sigma(1)\approx 1$, so that $\boldsymbol{z}_1$ is approximately standard Gaussian noise.
Let $\boldsymbol{z}_t$ denote a sample at time $t$, the definition of $\boldsymbol{z}_t$ based on the deterministic probability flow ODE \cite{ODE} can be written as: 
\begin{equation}
    \frac{\mathrm{d}\boldsymbol{z}_t}{\mathrm{d}t}
    \;=\; \mathbf{v}(\boldsymbol{z}_t, t)
        \;=\; \dot{\alpha}(t)\,\boldsymbol{z}_0 + \dot{\sigma}(t)\,\epsilon,
    \label{eq:ode}
\end{equation}
where $\mathbf{v}(\boldsymbol{z}_t, t)$ is the optimal velocity field that transports the marginal distribution from $q_0$ to $q_1$. The $\dot{\alpha}(t)$ and $\dot{\sigma}(t)$ denote time derivatives.
Then the conditional velocity field is parameterized with a neural network:$\mathbf{v}_\theta(\boldsymbol{z}_t, t, \boldsymbol{c})$, where $\boldsymbol{c} = \{ \boldsymbol{p}, \boldsymbol{m}, \boldsymbol{t} \}$ is the conditioning information. The model is trained with the standard flow-matching loss:
\begin{equation}
\mathcal{L}_{\text{img}} \;=\; \mathbb{E}_{\boldsymbol{z}_0, \boldsymbol{\epsilon}, t}  \bigl\| \mathbf{v}_\theta(\boldsymbol{z}_t, t, \boldsymbol{c}) - (\boldsymbol{\epsilon} - \boldsymbol{z}_0) \bigr\|_2^2.
\end{equation}

\textbf{Text Generation - Discrete Diffusion.}
In addition to image generation, our model also generates attribute-level structured texts. We define the clean text sequence as:
\begin{equation}
    \boldsymbol{y}_0 = (y_0^{(1)}, \dots, y_0^{(L)}) \in \{1,\dots,V\}^L,
\end{equation}
where $V$ is the vocabulary size and $L$ is the maximum sequence length.
Our goal is modeling conditional distribution $p_\theta(\boldsymbol{y}_0 \mid \boldsymbol{y}_t, \boldsymbol{c})$.
Following the discrete diffusion formulation in \cite{dualDiffusion}, the model uses a continuous-time Markov chain on the discrete state space:
\begin{equation}
    \mathcal{S} = \{1,\dots,V\} \cup \{<extra\_id0>\},
\end{equation}
where $<extra\_id0>$ indexes a special mask token.
For each token position, as $t$ increases, each position becomes masked with increasing probability, and at $t=1$ the sequence is totally masked.
Under the continuous-time limit, we define the masked language modeling objective:
\begin{equation}
\mathcal{L}_{\text{text}} 
= \mathbb{E} \left[ -\frac{1}{K} \sum_{i=1}^{K} \log \left[ \boldsymbol{y}_{\theta} \left( \boldsymbol{y}_{t_i}, \boldsymbol{c} \right) \cdot \boldsymbol{y}_0 \right] / t_i \right],
\end{equation}
where $K$ denotes the number of uniform time steps used for antithetic sampling to minimize the variance of the loss estimator.



\begin{figure*}[ht]
  \centering
  \setlength{\abovecaptionskip}{0.2cm} 
  \setlength{\belowcaptionskip}{-0.2cm}
  \includegraphics[width=1.0\textwidth]{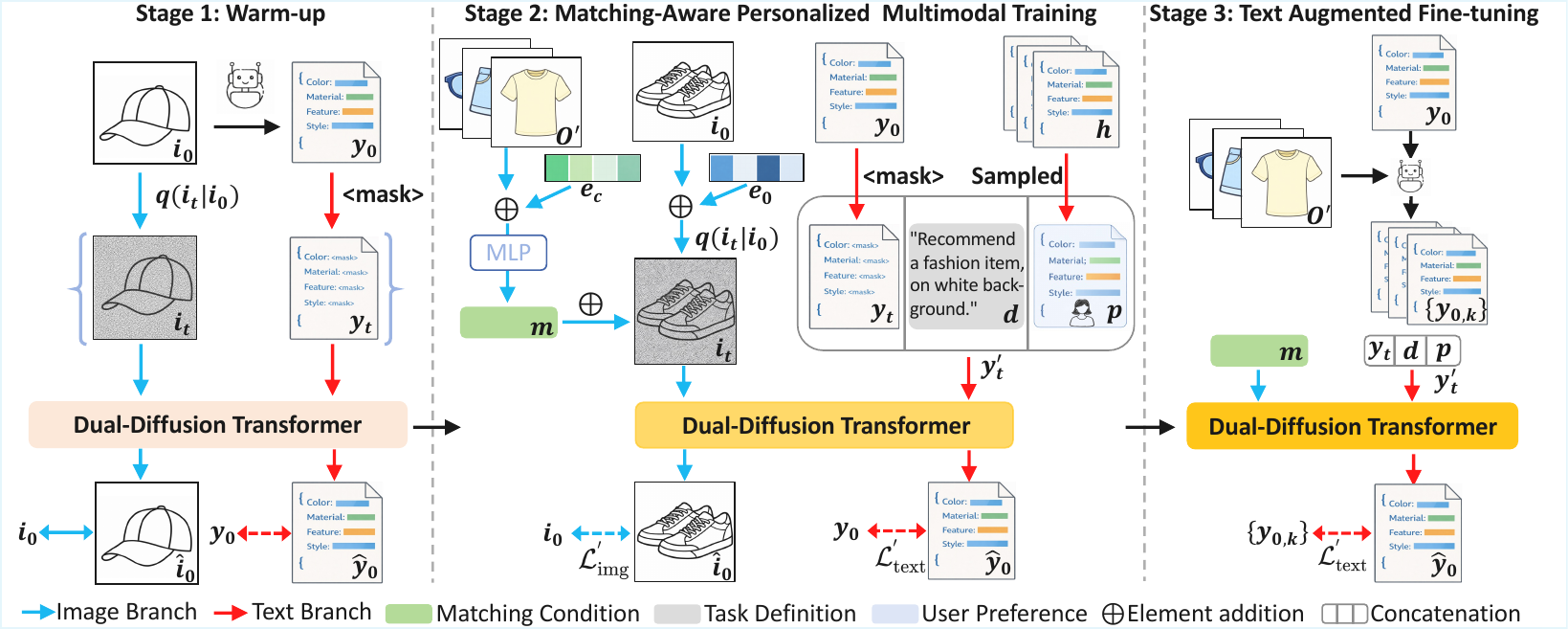} 
  \caption{
   The multi-stage training of DualFashion consists of warm-up, matching-aware personalized multimodal training, and text-augmented fine-tuning. The inference stage is omitted for clarity.
  }
  \label{fig:model}
\end{figure*}

\section{Our Approach}

In this section, we introduce multiple training stages, including warm-up, matching-aware personalized multimodal training, and text-augmented fine-tuning, followed by the inference stage.
Inspired by the Dual-Diffusion \cite{dualDiffusion}, our DualFashion is a Transformer-based model with image and text branches that leverage multimodal condition signals to model user behavior and generate multimodal outputs for explicit interpretability.

\subsection{Stage 1: Warm-up}


The warm-up stage adapts the general-domain Dual-Diffusion model to the fashion domain and performs cross-modal alignment using fashion image–caption pairs.
We define the caption $\boldsymbol{y} = {(a, V_a)}, {a \in A}$ with key set $A$ = \{Color, Material, Design features, Clothing Fashion Style\}
and encode the matching fashion item $\boldsymbol{i}_0$ into latent space $\boldsymbol{z}_0 = E(\boldsymbol{i}_0) \in \mathbb{R}^{C\times H\times W} $.
To be noted, the captions in large-scale image-text datasets are primarily composed of natural language. Consequently, when pre-training on fashion-specific datasets that pair images with structured attributes, models struggle to represent these structured semantics.
To address this, we explicitly supervise the structure learning by masking the value $V_a$ while keeping keys $a$ intact. This forces the model to reconstruct the missing values in a structured caption, thereby enhancing its capability to capture explicit attribute-level dependencies.

Specifically, in the text branch, given the ground-truth caption $\boldsymbol{y}_0$ and a binary label mask $\boldsymbol{M} \in \{0,1\}$ that marks the value tokens, we form the masked input:
\begin{equation}
    \boldsymbol{y}_t = \boldsymbol{y}_0 \odot (1-\boldsymbol{M}) + \boldsymbol{M} \odot \text{Mask}(\boldsymbol{y}_0,t),
    \label{eq:y_t}
\end{equation}
where the $\text{Mask}(\cdot,t)$ replaces masked positions with the mask token $<extra\_id0>$ according to the diffusion step $t$. The text diffusion loss is applied only on masked value positions:
\begin{equation}
    \mathcal{L}_{\text{text}} = -\sum_{j} m_j \log p_\theta(\boldsymbol{y}_{0,j} \mid \boldsymbol{y}_t, \boldsymbol{z}_0),
\end{equation}
where $\boldsymbol{z}_0$ is the image latent and $p_\theta$ is the transformer's conditional token distribution.

In the image branch, given the fashion item embedding $\boldsymbol{z}_0$, we inject Gaussian noise with a randomly sampled timestep $t$ and obtain noisy latent:
\begin{equation}
    \boldsymbol{z}_t = (1-\sigma_t)\boldsymbol{z}_0 + \sigma_t \boldsymbol{\epsilon}, \quad \boldsymbol{\epsilon}\sim\mathcal{N}(0,\mathbf{I}),
    \label{eq:z_t}
\end{equation}
where $\sigma_t$ is the noise scale determined by the scheduler. The transformer predicts noise with the caption embedding $\boldsymbol{y}_0$, and the image loss is defined as:
\begin{equation}
    \mathcal{L}_{\text{img}} \;=\; \mathbb{E}_{\boldsymbol{i}_0, \boldsymbol{\epsilon}, t}  \bigl\| \mathbf{v}_\theta(\boldsymbol{i}_t, t, \boldsymbol{y}_0) - (\epsilon - \boldsymbol{i}_0) \bigr\|_2^2.
\end{equation}
The total objective is a weighted sum: 
\begin{equation}
    \mathcal{L} = \mathcal{L}_{\text{img}} + \lambda_{\text{text}} \mathcal{L}_{\text{text}},
    \label{eq:L_joint}
\end{equation}
where $\lambda_{\text{text}}$ is the hyperparameter.
This joint loss let the model learns to generate images that are consistent with captions, while also generate structured texts that faithfully describe images.



\subsection{Stage 2: Matching-Aware Personalized Multimodal Training}
This stage aims to enhance the model’s matching ability by jointly generating structured textual descriptions and visual images under conditional guidance.
Specifically, we model two types of conditioning signals: (1) personalized preferences derived from user behaviors, and (2) mix-and-match patterns among fashion items.
Accordingly, the conditioning information includes user preference $\boldsymbol{p}$, matching condition $\boldsymbol{m}$, and task definition $\boldsymbol{d}$.

\textbf{$\boldsymbol{p}$: User Preference.} 
The user's interaction history reflects their specific preferences.
However, if the generation process only rely on image-level interactions, it will introduces the mismatch between user preference and generate item.
Because not all attributes in the interacted images $\boldsymbol{h}$ align with the user's true preferences.
For instance, a user may interact with an item solely for its design, disregarding its color.
To analyze user preference in the attribute-level, we firstly design template to extract the structured captions of item images.
Then we design Preference Weighted Sampling strategy to get textual user preference.

1) Structured Caption Extraction:
We define $A$ = \{Color, Material, Design features, Clothing Fashion Style\} is the set of attributes for fashion item. 
For each attribute $a \in A$, we we query a vision–language model \cite{Gemini} $g(\cdot)$ with structured prompt $p$ that specifies these attributes.
Given an fashion item image $\boldsymbol{i}$, the model outputs the attribute-level caption $V_a = g(\boldsymbol{i}, p, a), a \in A$.
We perform this structured caption extraction on all fashion items within the iFashion \cite{POG} and Polyvore-U \cite{Polyvore} datasets.

2) Preference Weighted Sampling:
For each attribute $a \in A$, there is a set of extracted elements $V_a$ and corresponding frequency score $f(v)$, which represents how often the fashion element appears in the interaction history.
A straight way to get the user preference is set the frequency threshold, but it may overlook sparse preferences. And if no frequency score reaches the threshold, it may lead to a loss of preference information.
So we propose the Preference Weighted Sampling, which sample fashion elements $V_a$ for each attribute $a \in A$ independently and the probability of fashion elements $V_a$ is proportional to its historical frequency score.
We define the probability $P(v)$ of sampling a specific element $v$ from $V_a$ for a given attribute $a$ as the normalized frequency score:
\begin{equation}
    P(v) = \frac{exp(f(v)/T)}{\sum_{v' \in V_a} exp(f(v')/T)},
\end{equation}
where $f(v)$ is the frequency score of element $v$, and $\sum exp(f(v')/T)$ is the sum of the exponentiated, temperature-scaled scores for elements $v'$ in the elements set $V_a$. 
The temperature $T$ controls the balance between exploiting known preference if $( T < 1)$ and exploring new ones if $( T > 1)$.
For each attribute $a \in A$, we independently sample N=10 times, remove duplicates, and obtain user preference.

\textbf{$\boldsymbol{m}$: Matching Condition.} 
For the target item $\boldsymbol{i}_0$, the generated result needs to be matching with the given incomplete outfit $\boldsymbol{O}^{'}$.
We extract the matching condition via the Stable Diffusion 3 autoencoder $E_{VAE}$ ~\cite{SDv3}, which normalizes each RGB image and maps it to a latent sample.
Given target latent $\boldsymbol{z}_0 \in \mathbb{R}^{C\times H\times W}$ and matching condition latent $\{\boldsymbol{z}_c^{(k)}\}_{k=1}^n$, we add learnable role embeddings $\boldsymbol{e}_0, \boldsymbol{e}_c \in \mathbb{R}^{C\times 1\times 1}$ to form: 
\begin{equation}
    \tilde{\boldsymbol{z}}_0 = \boldsymbol{z}_0 + \boldsymbol{e}_0, \quad 
    \tilde{\boldsymbol{z}}_c = \frac{1}{n}\sum_{k=1}^{n} \boldsymbol{z}_c^{(k)} + \boldsymbol{e}_c.
    \label{eq:latent_role_em}
\end{equation}
Then we project $\tilde{z}_c$ using a two-layer MLP equipped with LeakyReLU and dropout, and reshape the output back to get the matching condition $\boldsymbol{m} = \mathrm{MLP}(\tilde{z}_c)$.


\textbf{$\boldsymbol{d}$: Task Definition.} 
The task definition is defined as a template sentence: "Recommend a fashion {category} item, on white background."
This design injects the minimal semantic information (the item category) and a controlled visual signal (white background) to reduce spurious context.
We denote the template sentence as $s$ and its text encoder embedding is $\boldsymbol{d} = E_{\text{text}}(s)$, where the $E_{\text{text}}$ is T5 encoder.

\begin{table*}[h]
\centering
\setlength{\tabcolsep}{2pt}
\caption{Performance comparison between our method and various baselines. "Comp.", "Per." and "Div." denote compatibility, personalization and diversity, respectively. Bold indicates the best results while underline denotes the second best results.}
\vspace{-0.2cm}
\begin{tabular}{lccccccccc|ccccccccc}
\toprule
Dataset & \multicolumn{9}{c}{iFashion} & \multicolumn{9}{c}{Polyvore-U} \\
\cmidrule(lr){2-10} \cmidrule(lr){11-19} 
Task & \multicolumn{5}{c}{PFITB} & \multicolumn{4}{c}{GOR} &\multicolumn{5}{c}{PFITB} & \multicolumn{4}{c}{GOR} \\
\cmidrule(lr){2-6} \cmidrule(lr){7-10} \cmidrule(lr){11-15} \cmidrule(lr){16-19} 
Evaluation metric & IS  & IS-acc & Comp. & Per. & Div. & IS  & IS-acc & Per. & Div. & IS  & IS-acc & Comp. & Per. & Div. & IS  & IS-acc & Per. & Div.\\
\midrule
SD-v1.5~\cite{SD}* & 22.54  & 0.76 & 0.08 & 46.31 & \underline{0.56} & 23.20 & 0.77 &  46.45 & \underline{0.57} & 17.10  & 0.73 & 0.70  & 51.05 & \underline{0.55} & 16.95 & 0.73 & 50.99 & \underline{0.52}\\
SD-v2~\cite{SD}* & 21.66 & 0.71 & 0.04 & 46.60 & \textbf{0.62} & 22.19 & 0.74 &  46.60 & \textbf{0.59} & 14.83 & 0.68 & 0.60  & 51.29 & \textbf{0.60} & 14.88 & 0.67 & 51.23 & \textbf{0.65}\\
SD-v1.5~\cite{SD} & 26.76 & 0.83 & 0.46 & 53.16 & 0.50 & 26.90 & 0.84 &  53.24 & 0.49 & 17.12 & 0.72 & 0.75 & 58.20 & 0.50 & 17.24 & 0.72 & 58.16 & 0.47\\
SD-v2~\cite{SD} & 25.85 & 0.80 & 0.39 & 52.99 & 0.47 & 25.82 & 0.82 &  53.06 & 0.47 & 15.59 & 0.67 & 0.71  & 58.79  & 0.46 & 16.33 & 0.70 & 58.91 & 0.44\\
SD-naive~\cite{SD} & 25.45 & 0.80 & 0.36 & 52.95 & 0.39 & 25.43 & 0.81 &  52.95 & 0.40 & 15.45 & 0.66 & 0.73  & 59.24  & 0.41 & 15.48 & 0.67 & 59.12 & 0.40\\
ControlNet~\cite{controlnet} & 27.76 & 0.81 & 0.16 & 49.90  & 0.41 & 28.49  & 0.82 & 49.91  & 0.43 & 18.93 & 0.77 & 0.73  & 55.44  & 0.42 & 19.21 & 0.77 & 55.40 & 0.41\\
DiFashion~\cite{Difashion} & 29.99 & \underline{0.90} & 0.58 & 55.86 & 0.33  & 30.04  & \underline{0.90} & 55.54 & 0.33 & 19.67  & \underline{0.84} & 0.80  & 61.44  & 0.37 & 18.95  & 0.83 & 61.16 & 0.36\\
FashionDPO~\cite{FashionDPO}& \textbf{33.80} & \textbf{0.91} & \textbf{0.74} & \underline{60.39} & 0.36 & \textbf{32.37}  & \textbf{0.91} & \underline{59.98} & 0.35 & \underline{24.14} & \textbf{0.89} & \textbf{0.83} & \underline{64.67} & 0.38 & \textbf{24.93} & \underline{0.87} & \underline{64.79} & 0.36\\
\midrule
\rowcolor{gray!30} 
\textbf{DualFashion} & \underline{31.08} & 0.88 & \underline{0.69} & \textbf{65.17} & 0.38 & \underline{31.53}  & 0.89 & \textbf{64.05} & 0.39 & 
\textbf{25.30} & \textbf{0.89} & \underline{0.81} & \textbf{67.40} & 0.42 &
\underline{24.85} & \textbf{0.88} & \textbf{67.13} & 0.40 \\
\bottomrule
\end{tabular}
\label{tab:quantitative}
\end{table*}

In matching condition $\boldsymbol{m}$, both the target item $\boldsymbol{i}_0$ and incomplete outfit $\boldsymbol{O}^{'}$ come from the same dataset and are processed via a shared VAE encoder. Consequently, their latent representations lie in the same latent space and share similar marginal statistics, potentially causing the model to struggle in distinguishing these embeddings.
In our model, we introduce role embeddings $\boldsymbol{e}_0, \boldsymbol{e}_c$ to provide an explicit signal that distinguishes different semantic roles (target item and matching context).
We can get the latent distributions $\tilde{\boldsymbol{z}}_0$ and $\tilde{\boldsymbol{z}}_c$ from Eq. (\ref{eq:latent_role_em}) with means $\boldsymbol{\mu}_0$ and $\boldsymbol{\mu}_c$. And the mean gap between $\tilde{\boldsymbol{z}}_0$ and $\tilde{\boldsymbol{z}}_c$ is defined as:
\begin{equation}
    \Delta \boldsymbol{\mu} = \mathbb{E}[\tilde{\boldsymbol{z}_0}] - \mathbb{E}[\tilde{\boldsymbol{z}_c}] 
    = (\boldsymbol{\mu}_0 + \boldsymbol{e}_0) - (\boldsymbol{\mu}_c + \boldsymbol{e}_c) 
    = (\boldsymbol{\mu}_0 - \boldsymbol{\mu}_c) + (\boldsymbol{e}_0 - \boldsymbol{e}_c),
\end{equation}
where even when target and context latents largely overlap $\boldsymbol{\mu}_0\approx\boldsymbol{\mu}_c$, we still have $\Delta \boldsymbol{\mu}\approx(\boldsymbol{e}_0 - \boldsymbol{e}_c)\neq0$ that introduce mean shift that makes roles linearly decodable.

To fully leverage multimodal information, we feed user preference $\boldsymbol{p}$ and task definition $\boldsymbol{d}$ into the text branch, while providing the matching condition $\boldsymbol{m}$ through the image branch. 
We further optimize the model’s matching capability using a joint loss. Specifically, for the text loss, we adopt a masked text diffusion objective over caption tokens. At each step, we sample a masking rate and construct the masked caption $\boldsymbol{y}_t$ by replacing value tokens with the mask token as in Eq. (\ref{eq:y_t}). The text branch input $\boldsymbol{y}_t^{'}$ is formed by concatenating user preference and task definition with the masked caption: 
$ [\boldsymbol{y}_t;\boldsymbol{p};\boldsymbol{d}]$.
It is provided to attention blocks and attends to image branch input $\boldsymbol{m}$ at every attention layer, thereby allowing both textual and visual conditions to jointly guide the prediction of masked text tokens:
\begin{equation}
    \mathcal{L}_{\text{text}}^{'} = -\sum_{j} m_j \log p_\theta(\boldsymbol{y}_{0,j} \mid \boldsymbol{y}_t, \boldsymbol{p},\boldsymbol{m},\boldsymbol{d}).
    \label{eq:L_text}
\end{equation}
For the image loss, we use a flow‑matching reconstruction objective in latent space, where the noise latent $\boldsymbol{y}_t$ is initialized according to Eq. (\ref{eq:z_t}). 
We then inject the matching condition $\boldsymbol{m}$ into the noisy latent using a fixed mixing weight $\lambda_m$:
\begin{equation}
    \boldsymbol{z}_t^{'} = (1-\lambda_m)\boldsymbol{z}_t + \lambda_m \boldsymbol{m}.
\end{equation}
The image loss is calculated with the text branch condition $[\boldsymbol{p},\boldsymbol{d}]$:
\begin{equation}
    \mathcal{L}_{\text{img}}^{'} \;=\; \mathbb{E}_{\boldsymbol{z}_0, \boldsymbol{\epsilon}, t}  \bigl\| \mathbf{v}_\theta(\boldsymbol{z}_t^{'}, t, \boldsymbol{p},\boldsymbol{d}) - (\boldsymbol{\epsilon} - \boldsymbol{z}_0) \bigr\|_2^2.
\end{equation}
The joint training objective is defined as Eq. \ref{eq:L_joint}. 
To be noted, while the PFITB task aims to generate fashion item based on incomplete outfit, the GOR task generates a completed outfit without matching information.
Consequently, we employ a 50\% random drop rate for the matching condition during training. This strategy facilitates effective generalization from the PFITB to GOR task.

\subsection{Stage 3 - Text Augmented Fine-tuning.}
In the Dual-Diffusion Transformer, the visual and textual branches are jointly optimized during training, enabling the two modalities to mutually reinforce each other.
This joint optimization allows knowledge learned in the text branch transferred to the image branch, thereby enhancing visual generation.
We augment the textual data using the advanced LLM Gemini \cite{Gemini}, which possesses extensive fashion-domain knowledge.
Specifically, we generate diverse yet matching fashion item texts for incomplete outfits $\boldsymbol{O}^{'}$, enriching the textual ground truth with high-quality, diversity descriptions. 
During the fine-tuning stage, the matching information $\boldsymbol{m}$ is provided to the image branch, while the text branch takes the constructed input $\boldsymbol{y}_t^{'}$. 
Leveraging the augmented captions, we fine-tune the model using only the text loss defined in Eq.~(\ref{eq:L_text}), explicitly focusing the optimization on textual representations.
Compared to image-based reinforcement learning and fine-tuning, text-based fine-tuning is significantly more efficient, as it avoids costly diffusion image sampling and high-dimensional visual optimization.

\subsection{Inference Stage}

We introduce the sampling process for fashion item texts and images in the PFITB task, where the GOR task can be viewed as performing $n$ rounds of sampling:
in the first round, no matching condition is provided, and in the $i$-th round, the images generated in the previous $i-1$ rounds are used as the matching condition. After $n$ rounds, this process yields an outfit composed of $n$ fashion items.

\textbf{Textual Output.} We first construct a structured caption template $\boldsymbol{y}_t$ whose attribute values are masked as in Eq. (\ref{eq:y_t}).
The sampler then performs discrete diffusion over the masked tokens, iteratively predicting values to approximate 
$p_\theta(\boldsymbol{y}_0 \mid \boldsymbol{y}_t, \boldsymbol{p}, \boldsymbol{m}, \boldsymbol{d})$.
After a fixed number of diffusion steps, the resulting tokens are decoded into a fashion item's structured text. 

\textbf{Visual Output.} We first construct the text condition sequence by concatenating the user preference $\boldsymbol{p}$ and task definition $\boldsymbol{d}$, denoted as $[\boldsymbol{p}, \boldsymbol{d}]$. 
For the image branch, the latent representation $\boldsymbol{z}_t^{'}$ is initialized from empty image with standard Gaussian noise.
To ensure the generated fashion item is matching with the incomplete outfit, we fuse the $\boldsymbol{z}_t^{'}$ with matching information $\boldsymbol{m}$. 
Subsequently, we employ the Classifier-Free Guidance (CFG) \cite{CFG} to sample fashion item image: 
\begin{equation}
\begin{aligned}
\hat{v}_t ={}& v_\theta\!\left(\boldsymbol{z}_t^{'}, \varnothing, \varnothing, \varnothing, t \right) \\
&+ s_d \cdot \big[
    v_\theta\!\left(\boldsymbol{z}_t^{'}, \boldsymbol{d}, \varnothing, \varnothing, t \right)
    - v_\theta\!\left(\boldsymbol{z}_t^{'}, \varnothing, \varnothing, \varnothing, t \right) 
\big] \\
&+ s_m \cdot \big[
    v_\theta\!\left(\boldsymbol{z}_t^{'}, \boldsymbol{d}, \boldsymbol{m}, \varnothing, t \right)
    - v_\theta\!\left(\boldsymbol{z}_t^{'}, \boldsymbol{t}, \varnothing, \varnothing, t \right)
\big] \\
&+ s_p \cdot \big[
    v_\theta\!\left(\boldsymbol{z}_t^{'}, \boldsymbol{d}, \boldsymbol{m}, \boldsymbol{p}, t \right)
    - v_\theta\!\left(\boldsymbol{z}_t^{'}, \boldsymbol{t}, \boldsymbol{m}, \varnothing, t \right)
\big],
\end{aligned}
\end{equation}
where $s_d, s_m, s_p$ are hyperparameters that control the influence of each condition.
By linearly interpolating between the conditional and unconditional noise predictions, we guide the generation towards high fidelity, ensuring both outfit matching and adherence to user preference. Finally, the denoised latent $\hat{\boldsymbol{z}}_0$ is decoded by the VAE decoder to produce the fashion item image. 






\section{Experiments}

We conduct experiments on the iFashion and Polyvore-U datasets, aimming to answer the following research questions (RQs):
\begin{itemize}[leftmargin=1.2em]
\item \textbf{RQ1:} The effectiveness of modeling user behavior with multimodal information.
Compared with other generative baselines, does the model show improvements in quantitative metrics?
\item \textbf{RQ2:} What are the effects of the warm-up stage, learnable embeddings, loss design, and text-augmented fine-tuning in our proposed dual-diffusional architecture?

\item \textbf{RQ3:} Does our method achieve better qualitative results than SOTA fashion generative recommendation models?


\end{itemize}

\subsection{Experimental Settings}

\subsubsection{\textbf{Baselines}}

For the PFITB and GOR tasks, we compare our model with the following baselines:
\textbf{1) SD-v1.5}~\cite{SD}: It's a latent space diffusion model. In the model names, with "*" indicates a pre-trained model, while without "*" indicates that the model has been fine-tuned on the fashion dataset.
\textbf{2) SD-v2}: It's an upgraded version of SD-v1.5. The same naming convention.
\textbf{3) SD-naive}: It's a fine-tuned model based on SD-v2, where concatenate mutual influence and history condition as condition.
\textbf{4) ControlNet}~\cite{controlnet}: It's an extension model based on SD, which introduces additional conditional inputs.
\textbf{5) DiFashion}~\cite{Difashion}: It's the SOTA generative recommendation model based on SD-v2, which utilizes a parallel diffusion process to generate fashion items.
\textbf{6) FashionDPO} ~\cite{FashionDPO}: It's a multi-expert feedback framework that fine-tunes the fashion outfit generation model using Direct Preference Optimization ~\cite{D3PO}.

\begin{table*}[t]
\centering
\small
\setlength{\tabcolsep}{2pt}
\caption{Ablation study about the impact of warm-up, learnable embeddings, loss design and text augmented fine-tuning on fashion item text generation. Bold indicates the best results while underline denotes the second best results.}
\vspace{-0.2cm}
\label{tab:ablation_others}
\begin{tabular}{lccccccccccccccc}
\toprule
 & \multicolumn{5}{c}{Comp.} 
 & \multicolumn{5}{c}{Per.} 
 & \multicolumn{5}{c}{Div.} \\
 
\cmidrule(lr){2-6}
\cmidrule(lr){7-11}
\cmidrule(lr){12-16}

Method
& Color & Material  & Features  & Style & Ave.
& Color & Material  & Features  & Style & Ave.
& Color & Material  & Features  & Style & Ave.\\
\midrule
w/o warm-up
& 0.70 & 0.71 & 0.33 & 0.42 & 0.54
& 0.49 & 0.65 & 0.50 & 0.67 & 0.58
& 3.31 & 2.25 & 4.90 & 2.82 & 3.32 \\

w/o learnable embeddings
& 0.84 & 0.70 & 0.37 & 0.55 & 0.62
& \underline{0.56} & \underline{0.72} & 0.53 & 0.74 & 0.64
& 3.36 & 2.28 & 4.95 & 2.81 & 3.35 \\

\midrule
only image loss 
& 0.63 & 0.68 & 0.30 & 0.34 & 0.49
& 0.50 & 0.57 & 0.34 & 0.62 & 0.51
& \textbf{4.03} & \textbf{4.95} & \underline{5.51} & \textbf{5.38} & \textbf{4.97}\\ 

only text loss   
& \underline{0.89} & 0.81 & \underline{0.40} & 0.56 & 0.66
& \textbf{0.57} & \textbf{0.74} & \textbf{0.60} & \textbf{0.84} & \textbf{0.69}
& 3.21 & 2.72 & \textbf{5.52} & 2.48 & 3.48\\

image and text loss 
& \textbf{0.91}  & \textbf{0.84}  & 0.39  & \underline{0.59}  & \underline{0.68}
& \textbf{0.57} & \textbf{0.74} & 0.55& 0.80 & 0.67
& 3.38 & 2.63 & \textbf{5.52} & 3.03 & 3.64 \\

\midrule


w/ text augmented fine-tuning \textbf{(Ours)}
& \textbf{0.91}  & \underline{0.83}  & \textbf{0.44}  & \textbf{0.71}  & \textbf{0.72}
& \textbf{0.57} & \textbf{0.74} & \underline{0.57} & \underline{0.82} & \underline{0.68}
& \underline{3.47} & \underline{3.64} & \textbf{5.52} & \underline{4.36} & \underline{4.25} \\
\bottomrule
\label{tab:ablation_text}
\end{tabular}
\vspace{-0.5cm}
\end{table*}

\subsubsection{\textbf{Datasets}}

Our experiments are built on two existing datasets, iFashion\cite{POG} and Polyvore-U \cite{Polyvore}, which contains user, fashion item, outfit and user-item interaction history.
We further augment these datasets by extracting structured captions and user preferences. 
The iFashion dataset contains 50 pre-defined common fashion categories.
For each outfit, there are four mutually matching  fashion items. 
We extract all 344186 fashion item structured captions and sample user preferences from multiple items.
For the Polyvore-U dataset, we use the classifier Inception-V3 \cite{Inception-V3} fine-tuned on the iFashion dataset to perform image recognition and classification on the fashion items, and apply the same procedure to extract structured captions and sample user preferences.

\begin{table}[t]
\setlength{\tabcolsep}{2pt}
\centering
\caption{Ablation study on fashion item image generation quality, compatibility, personalization and diversity.}
\vspace{-0.2cm}
\begin{tabular}{lccccc}
\hline
Method & IS & IS-acc & Comp. & Per. & Div. \\
\hline
w/o warm-up & 28.77 & 0.80 & 0.62 & 62.68 & 0.31 \\
w/o learnable embeddings & 30.23 & 0.85 & \underline{0.64} & 62.93 & 0.31 \\
\midrule
only image loss & 30.29 & \textbf{0.93} & \textbf{0.71} & 63.41 & \underline{0.36}\\
only text loss & 26.57 & 0.84 & 0.65 & \textbf{65.86} & 0.33 \\
image and text loss & \underline{30.61} & 0.87 & \underline{0.69} & \underline{64.48} & 0.34 \\
\midrule
\makecell[l]{w/ text augmented fine-tuning \\ \textbf{(Ours)}}  & \textbf{31.08} & \underline{0.88} & \textbf{0.71} & \underline{65.17} & \textbf{0.38} \\
\hline
\end{tabular}
\label{tab:ablation_image}
\vspace{-0.4cm}
\end{table}

\subsubsection{\textbf{Evaluation Metrics}}

To evaluate the generated fashion item images, we employ a list of quantitative evaluation metrics, covering four major evaluating perspectives: 
\textbf{1) Quality}: We use Inception Score (IS) to evaluate the quality of the generated images. Additionally, we use IS-Accuracy (IS-acc) to further assess whether the generated images are correct regarding its category-level semantics. 
\textbf{2) Compatibility (Comp.)}: We follow the previous work~\cite{Difashion} and use the discriminator in OutfitGAN~\cite{OutfitGAN} to calculate the compatibility.
\textbf{3) Personalization (Per.)}: We use the foundation model CLIP~\cite{CLIP} to extract the image embeddings of the interacted items, then calculate the cosine similarity.
\textbf{4) Diversity (Div.)}: We use LPIPS to compute the diversity of the generated images between each pair of multiple generated images within each outfit.

Similarly, we design four evaluation perspectives for the generated fashion item texts and evaluate models' performance on four attributes $A$:
\textbf{1) Align}: We use the CLIP score to evaluate whether the generated texts are aligning with the generated images.
\textbf{2) Compatibility (Comp.)}: We use the advanced LLM Gemini to calculate the compatibility between generated texts and extracted incomplete outfit items' captions.
\textbf{3) Personalization (Per.)}: We calculate average CLIP score between the generated texts and sampled user preference.
\textbf{4) Diversity (Div.)}: We use the Semantic Distance (S.D.) and attribute entropy (A.E.) to measure the diversity of the generation distribution.

\subsubsection{\textbf{Implementation Details}} 
We use pre-trained Dual-Diffusion \cite{dualDiffusion} model as backbone. 
In stage 1 - Warm-up, we construct image–text pairs from fashion item images and their extracted captions, and fine-tune the model for 20,000 steps.
We set learning rate $3\times10^{-5}$ and weight decay $1\times10^{-2}$.
In stage 2 -  Matching-Aware Personalized Multimodal Training, the matching condition is fused with initialized noisy latent using a matching scale of $\lambda_m=0.1$.
When sampling user preference, we sample each attribute ten times and remove duplicate elements.
The training objective is the sum of image flow‑matching loss and text diffusion loss weighted by $\lambda_{\text{text}}=0.2$.
In stage 3 - Text Augmented Fine-tuning, we use Gemini to augment captions five times for each outfit, and finetune our model using only the text diffusion loss.
In the inference stage, we use the CFG to control these conditions and set the hyperparameters $s_d=8, s_m=7, s_p=8$.
All our experiments are conducted on L40s GPUs with 48 GB of memory.

\subsection{Quantitative Results}

In this section, we compare DualFahsion with baselines across visual and textual metrics, and validate the effectiveness of warm-up, learnable embeddings, loss design and text augmented fine-tuning through ablation study.

\subsubsection{\textbf{Overall Performance Comparison (RQ1)}}
To evaluate the effectiveness of modeling user behavior and model's generation ability, we compare models' performance on different datasets and tasks.
As shown in Tab. \ref{tab:quantitative}, our proposed DualFashion yields slightly lower scores than FashionDPO in terms of IS, IS-acc, and Comp. However, FashionDPO achieves this at the cost of high computational complexity. Compared to DiFashion, our method still demonstrates significant improvements across these metrics.
Notably, our DualFashion shows a significant improvement in the Personalization score (Per.), this indicates that our model is particularly effective at modeling user behavior and generating items that align with individual user preference.
In the Diversity score (Div.), the results show that pre-trained Stable Diffusion models (SD*) exhibit the highest Diversity scores across both datasets. However, this stems from randomness and lack of constraints, resulting in outputs that are visually distinct but incompatible and irrelevant to user preferences. 
Upon fine-tuning, the Diversity score declines as models tend to converge towards safe and frequent patterns within the data distribution to maximize compatibility, thereby suppressing diversity.
In contrast, our approach enhances diversity by leveraging user behavior modeling and effectively transferring outside fashion knowledge into the model.

\begin{figure}[t]
    \centering
    \setlength{\abovecaptionskip}{0.2cm} 
    \setlength{\belowcaptionskip}{-0.4cm} 
    \includegraphics[width=0.48\textwidth]{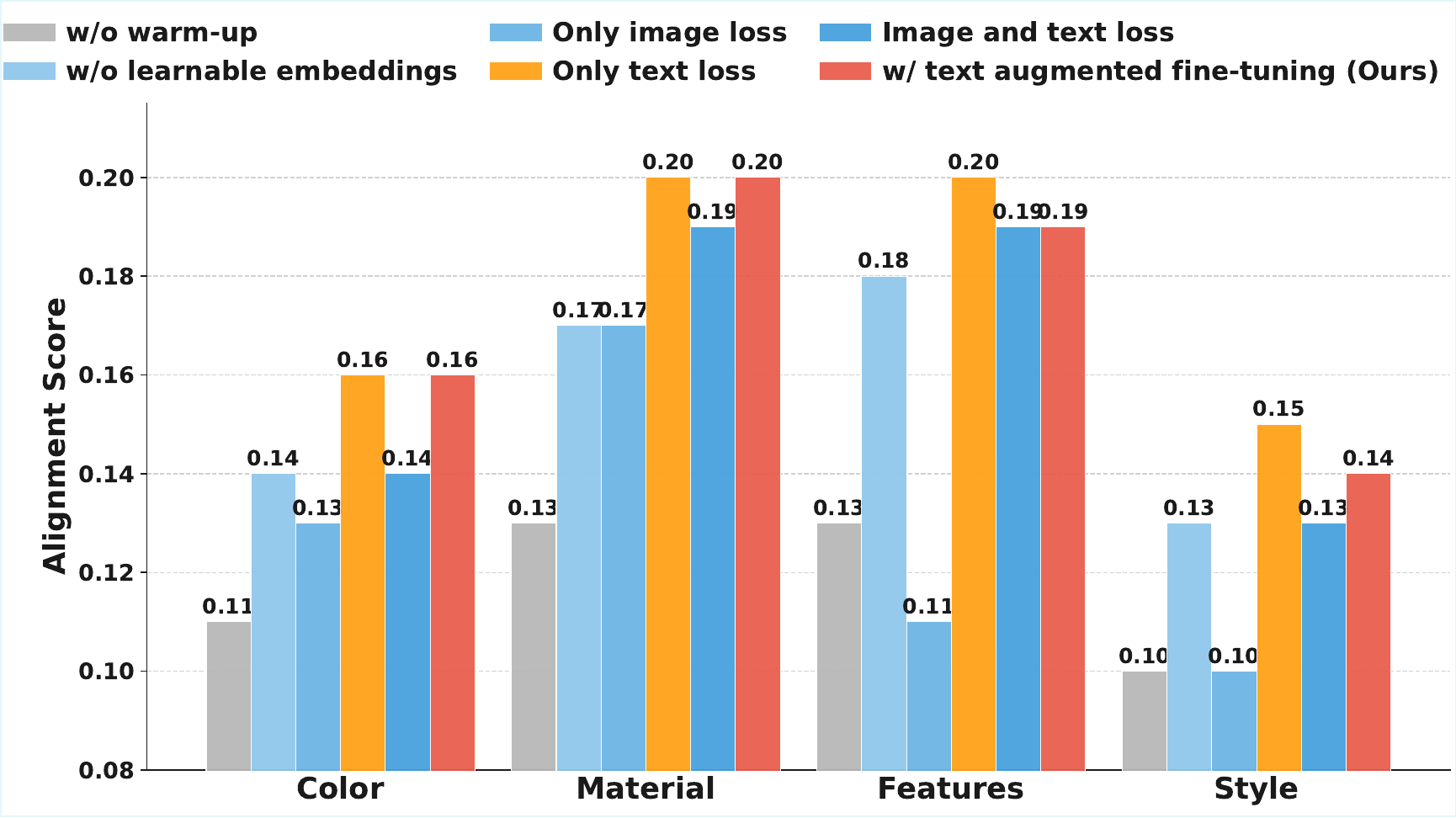} 
    \caption{Ablation study about the alignment between fashion item image and text generation.}
    \label{fig:alignment}
\end{figure}

\begin{figure*}[t]
    \centering
    \setlength{\abovecaptionskip}{0.1cm} 
    \setlength{\belowcaptionskip}{-0.2cm} 
    \includegraphics[width=1.0\textwidth]{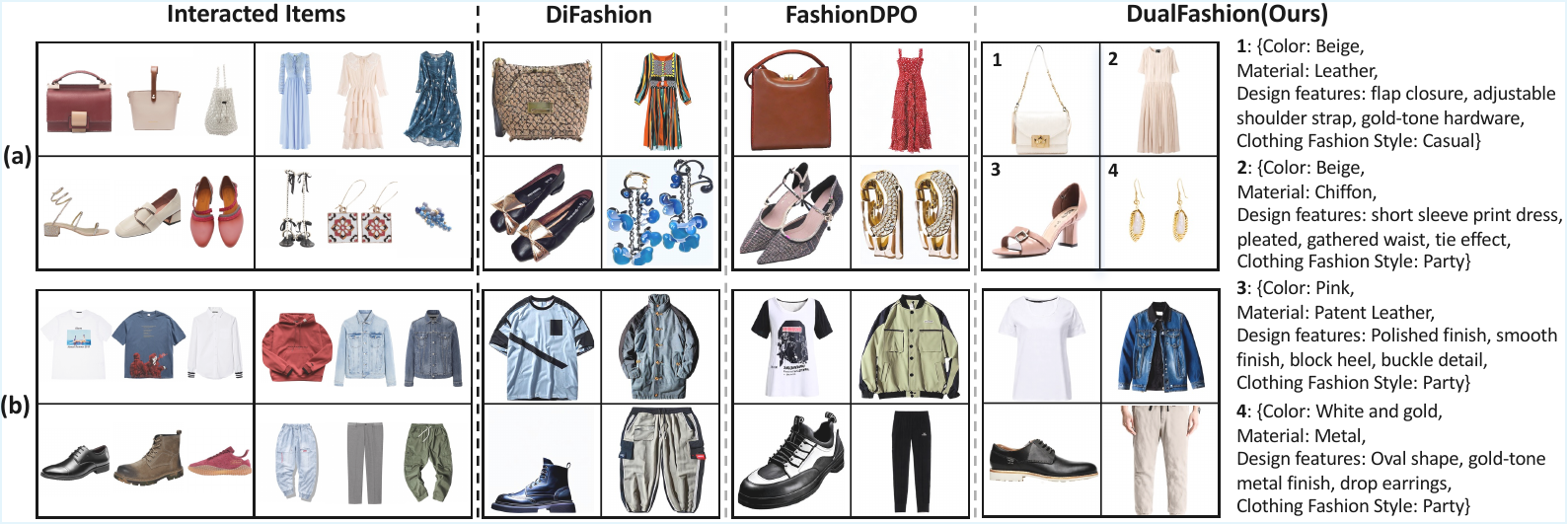} 
    \caption{Model-wise comparison of different models’ generative capabilities on the GOR task. Our DualFashion generates fashion item images and texts, producing outfits with higher compatibility while better aligning with user preferences.}
    \label{fig:gor}
\end{figure*}

\subsubsection{\textbf{Ablation Study (RQ2)}}

We conduct ablation studies on the iFashion dataset, evaluating performance across three dimensions: text generation, image generation and image-text alignment.

\textbf{Fashion item text generation.}
As shown in Tab. \ref{tab:ablation_text}, without warm-up results in bad performance in all metrics, while removing learnable embeddings has more impact on compatibility. This indicates that learnable embeddings help the model effectively distinguish the target item from the incomplete outfit.
Training with only image loss results in inferior compatibility and higher diversity score, indicating that image-only supervision fails to ensure textual consistency while producing less controlled outputs.
In contrast, optimizing with only text loss improves personalization performance, demonstrates that the text loss is more effective for personalization modeling.
Finally, incorporating text augmented fine-tuning significantly improve the compatibility and diversity score, suggesting that enriched fashion matching knowledge enhances matching ability while encouraging more diverse text generation.

\textbf{Fashion item image generation.} 
As shown in Tab. \ref{tab:ablation_image}, the complete method with text augmented fine-tuning achieves the most comprehensive performance. 
With only image loss, the model achieves significant improvements in IS-acc and Comp. scores, which demonstrates that only image supervision encourages the model to prioritize visual fidelity and compatibility.
In contrast, optimizing with only text loss yields higher personalization performance.
Therefore, it is necessary to balance the two objectives to jointly capture visual compatibility and personalization.

\textbf{Image-text alignment.} Fig. \ref{fig:alignment} shows the alignment performance across four attribute dimensions.
Without the warm-up degrades performance, it shows that this stage aligns visual appearances with structured textual captions.
For loss design, training with only image loss results in poor alignment, suggesting that image-only supervision is insufficient for learning attribute-level semantics.
In contrast, optimizing with only text loss achieves the best performance, as it directly enforces semantic consistency in text generation.
Moreover, compare with the joint image and text loss, text augmented fine-tuning under text-only loss improves the score, demonstrating the effectiveness of text loss in modality alignment.

%







\begin{figure}[t]
    \centering
    \setlength{\abovecaptionskip}{0.2cm} 
    \setlength{\belowcaptionskip}{-0.4cm} 
    \includegraphics[width=0.48\textwidth]{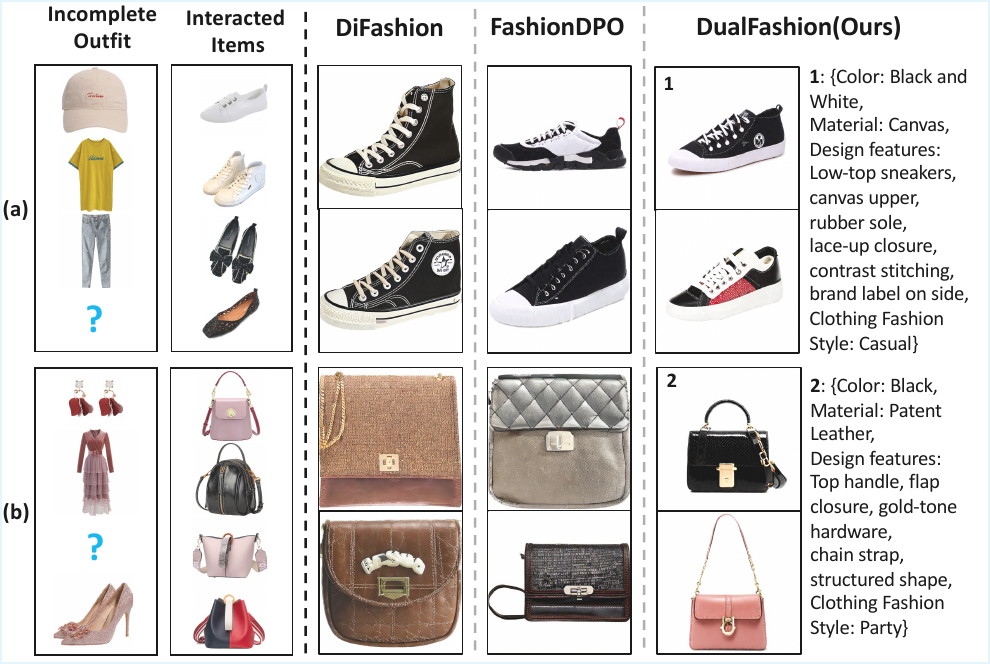} 
    \caption{Comparison on the PFITB task. Two generated images per model are shown for each incomplete outfit.}
    \label{fig:pfitb} 
\end{figure}



\begin{figure*}[t]
    \centering
    \captionsetup[subfigure]{labelformat=simple}
    \setlength{\abovecaptionskip}{0.3cm} 
    \setlength{\belowcaptionskip}{-0.2cm} 
    \begin{subfigure}[b]{0.38\textwidth}
        \centering
        \includegraphics[width=\linewidth]{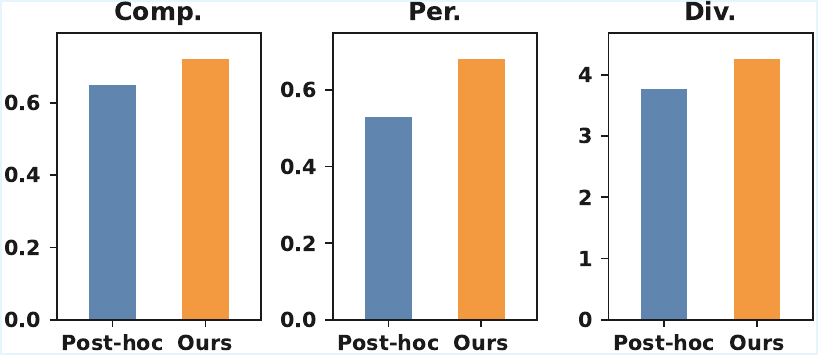}
        \caption{Comparison of interpretability ability.}
        \label{fig:modelStudy_interpretability}
    \end{subfigure}
    \hfill 
    \begin{subfigure}[b]{0.60\textwidth}
        \centering
        \includegraphics[width=\linewidth]{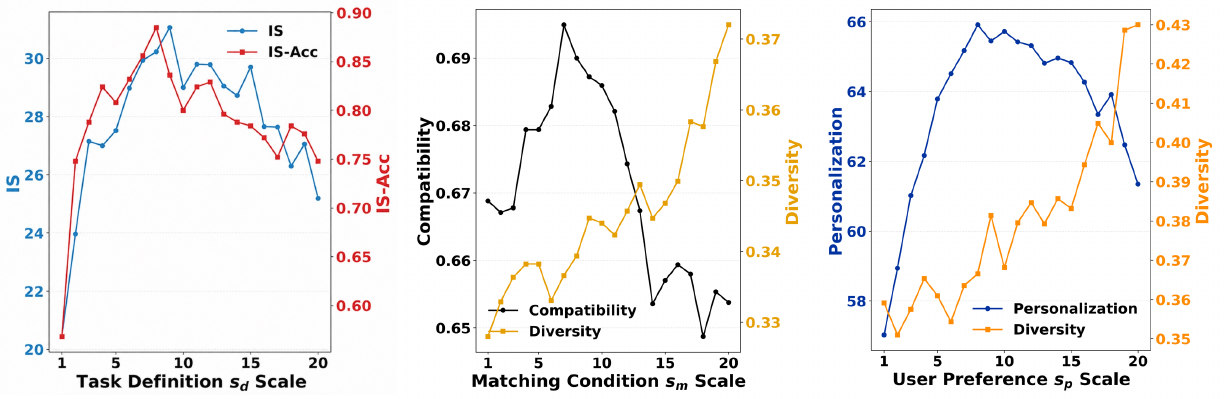}
        \caption{Effect of inference hyperparameters.}
        \label{fig:modelstudy_hyperparameters}
    \end{subfigure}
    \caption{Experimental Evaluation. (a) Comparison of interpretability ability between our model architecture and post-hoc caption generation. (b) Effect of inference hyperparameters $s_d$, $s_m$, and $s_p$ on model performance.}
    \label{fig:analysis_combined}
\end{figure*}

\begin{figure}[t]
    \centering
    \setlength{\abovecaptionskip}{0.1cm} 
    \setlength{\belowcaptionskip}{-0.4cm} 
    \includegraphics[width=0.48\textwidth]{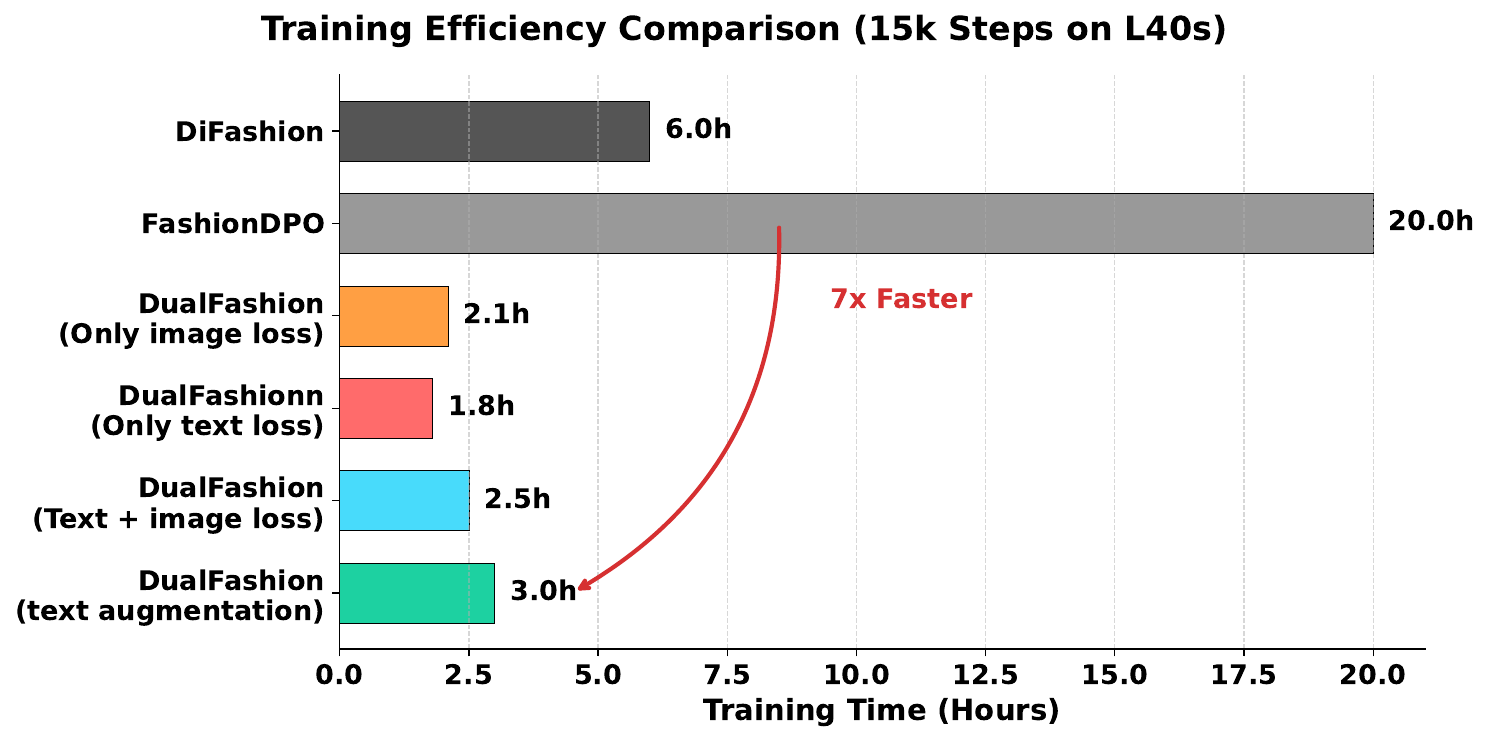} 
    \caption{Time cost analysis of the baseline and our DualFashion under identical training timesteps and hardware.}
    \label{fig:modelstudy_timecost}
\end{figure}

\subsection{Qualitative Results (RQ3)}
We compare qualitative results on GOR and PFITB tasks and analyze them from three perspectives: image generation quality, compatibility, and personalization.

\textbf{GOR task.} As shown in Fig. \ref{fig:gor}, given the same interacted items, all models generate visually plausible items; however, DiFashion in group (a) produces pants with distorted proportions, while 
FashionDPO in group (b) generates earrings with incorrect structural details.
In terms of outfit compatibility, DiFashion and FashionDPO may generate items that match individual pieces but lack global coherence. In contrast, DualFashion generates outfits with consistent color palettes and materials, \eg, beige dresses paired with gold-tone earrings or patent leather shoes that match the overall style.
Regarding personalization, DualFashion better reflects user preferences observed from interacted items, for instance, it generates patent leather shoes, pleated dresses, and bags with gold-tone hardware that align with the user’s demonstrated taste.

\textbf{PFITB task.} 
In terms of generation quality, DiFashion in group (b) produces bags with less structural details, whereas DualFashion maintains clean silhouettes and realistic materials, \eg, well-shaped shoes and leather handbags with clear edges.
In terms of compatibility, DiFashion and FashionDPO sometimes focus on local similarity but fail to ensure outfit coherence.
In contrast, DualFashion generates items that better complement the incomplete outfit, \eg, low-top canvas sneakers that match casual style or structured leather handbags that align with party style dress.
As for personalization, DualFashion generates items that are more strongly aligned with user preferences than the baselines, such as black-and-white casual sneakers or patent leather handbags with gold-tone hardware.


\subsection{Model Study}

\subsubsection{\textbf{Caption Interpretability}}

Different from prior fashion generative recommendation works, our model architecture is able to generate fashion item texts, providing clear interpretability.
As shown in Fig. \ref{fig:gor} and Fig. \ref{fig:pfitb}, the generated texts effectively capture key characteristics of the fashion items.
While for prior works, we can firstly generates the fashion item image, and then applies image-to-text model to generate the post-hoc caption.
To demonstrate the necessity of multimodal output, we compare the texts generated by our dual-diffusional architecture with post-hoc captions in terms of textual compatibility, personalization and diversity. 
As shown in Fig. \ref{fig:modelStudy_interpretability}, the personalization score has significant improvement.
We attribute this gain to the inherent difference between modalities: images represent low-level pixel information, often containing visual details, whereas texts represent high-level semantic information that explicitly defines user preferences.
The post-hoc approach loses information by inferring semantics from pixels, while our method directly decodes user preferences into text, preserving the integrity of the semantic intent.

\subsubsection{\textbf{Time Cost Analysis}}
We compare the training time costs of different models for the same number of training timesteps.
As shown in Fig. \ref{fig:modelstudy_timecost}, our DualFashion is approximately 7$\times$ faster than FashionDPO, indicating that text augmented fine-tuning at the textual level is significantly more computationally efficient than at the image level.
Moreover, by replacing the U-Net-based Diffusion with transformer-based Dual-Diffusion, our architecture significantly reduces the overall training time.
In addition, predicting only masked captions and optimizing the text loss incurs the lowest computational cost among all settings.


\subsubsection{\textbf{Hyper-parameter Analysis}}

We use the hyperparameters $s_d, s_m, s_p$ to control the three conditions User Preference $\boldsymbol{d}$, Matching Condition $\boldsymbol{m}$, Task Definition $\boldsymbol{p}$ in the inference stage.
We explore the model’s performance under different parameter settings, focusing on the two most relevant metrics.
As shown in Fig. \ref{fig:modelstudy_hyperparameters}, we can see that with the task definition scale $s_d$ increases, both the IS score and IS-Acc score first improve; however, when the value of $s_d$ becomes too large, the performance on these metrics degrades, suggesting that it may limit model flexibility and influence generation quality.
For the matching condition scale $s_m$ and user preference scale $s_p$, an appropriate value of $s_m$ and $s_p$ achieve the highest performance, while further increasing $s_m$ and $s_p$ gradually reduce compatibility and personalization but consistently improves diversity.
This indicates that overly strong matching and user preference constraints may reduce the model’s ability to produce compatible and personalized outputs, but encourage more diverse generations.
Overall, these results highlight the importance of hyperparameters $s_d$, $s_m$, and $s_p$ in balancing image quality, compatibility, personalization, and diversity during the inference stage.

\section{Conclusion And Future Work}
In this paper, we identified two limitations in existing image-centric generative recommendation methods: inefficient behavior modeling and a lack of model explainability.
To address these limitations, we propose DualFashion, a dual-diffusional architecture that jointly models user behavior and provide interpretable multimodal output.
Experiments on the iFashion and Polyvore-U datasets demonstrate that DualFashion more effectively models user behavior while delivering both visually compatible images and interpretable textual captions.
Moreover, by augmenting captions, our DualFashion achieves  more efficient training and improved generation diversity.

In future work, we plan to explore alternative multimodal architectures to better understand their strengths.
We will further explore how to integrate collaborative modeling of user–item interaction signals into image-based generative recommendation, and analyze how these signals contribute to preference learning.
Furthermore, we will investigate how interaction IDs can be effectively modeled to support fine-grained user behavior modeling.

\begin{acks}
This research was supported by the Singapore Ministry of Education (MOE) Academic Research Fund (AcRF) Tier 1 grant.
\end{acks}

\newpage
\bibliographystyle{ACM-Reference-Format}
\bibliography{main}

\newpage
\appendix

\end{document}